\documentclass[
10pt,
twocolumn,
aps,
prb,
amsmath,amssymb,
showpacs,
superscriptaddress,floatfix
]{revtex4-2}

\usepackage{color}
\usepackage{graphicx}% Include figure files
\usepackage{dcolumn}% Align table columns on decimal point
\usepackage{bm}% bold math
\begin{document}

%\preprint{APS/123-QED}

\title{The self-organized vacancy order in Pr$_9$Ge$_{16}$ }% Force line breaks with \\

\author{Jayashani S. T. Wickramasinghe}
\affiliation{Department of Physics, Missouri University of Science and Technology, Rolla, Missouri 65409, USA}

\author{Melissa G. Anderson}
\affiliation{Department of Chemistry and Biochemistry, Baylor University, Waco, Texas 76798, USA}

\author{Kelci Graville}
\affiliation{Department of Physics, Missouri University of Science and Technology, Rolla, Missouri 65409, USA}

\author{Gregory T. McCandless}
\affiliation{Department of Chemistry and Biochemistry, Baylor University, Waco, Texas 76798, USA}

\author{Zachary J. Morgan}
\affiliation{Neutron Scattering Division, Oak Ridge National Lab, Oak Ridge, Tennessee 37831, USA}

\author{Brianna R. Billingsley}
\author{Tai Kong}
\affiliation{Department of Physics, University of Arizona, Tucson, Arizona 85721, USA}

\author{Hyunsoo Kim}
\affiliation{Department of Physics, Missouri University of Science and Technology, Rolla, Missouri 65409, USA}
\affiliation{Materials Research Center, Missouri University of Science and Technology, Rolla, Missouri 65409, USA}

\author{Aleksandr V. Chernatynskiy}
\affiliation{Department of Physics, Missouri University of Science and Technology, Rolla, Missouri 65409, USA}

\author{Simon G. Mitchell}
\author{Liang Wu}
\affiliation{Department of Physics and Astronomy, University of Pennsylvania, Philadelphia, Pennsylvania 19104, USA}

\author{Julia Y. Chan}
\affiliation{Department of Chemistry and Biochemistry, Baylor University, Waco, Texas 76798, USA}

\author{Feng Ye}
\affiliation{Neutron Scattering Division, Oak Ridge National Lab, Oak Ridge, Tennessee 37831, USA}

\author{Halyna Hodovanets}
\affiliation{Department of Physics, Missouri University of Science and Technology, Rolla, Missouri 65409, USA}
\affiliation{Materials Research Center, Missouri University of Science and Technology, Rolla, Missouri 65409, USA}
\email{halyna.hodovanets@mst.edu}

\begin{abstract}
In this work, we report the discovery of a new crystal structure on the Ge‑rich side of the Pr–Ge binary phase diagram. Using a high‑temperature flux technique, we grew single crystals of Pr$_9$Ge$_{16}$, which adopt a previously unreported orthorhombic $Fdd$2 structure type featuring ordered Ge vacancies. We present the anisotropic magnetic properties and identify the crystallographic $b$ axis perpendicular to the crystal plane as the magnetic easy axis. Temperature‑dependent resistivity measurements reveal metallic behavior with a distinct anomaly at $T_C$ = 14.3 K. Hall resistivity data indicate that electron-like carriers
dominate, with a carrier concentration on the order of 10$^{27}$ m$^{-3}$. The magnetic order is readily suppressed by a magnetic field of approximately 0.4 T applied along the easy $b$ axis.
\end{abstract}

\keywords{Single crystals, flux growth, crystal structure, magnetic anisotropy, resistivity, Hall resistivity}%Use showkeys class option if keyword
                              %display desired
\maketitle

%\tableofcontents

\section{Introduction}

Rare earth digermanides, LnGe$_2$, (Ln-lanthanide) and their non-stoichiometric counterparts, LnGe$_{2-x}$, have been the focus of extensive study due to their rich variety of crystal structures and the intriguing physical properties they exhibit. Partial or complete ordering of vacancies in these compounds gives rise to a wide range of superstructures and leads to diverse electronic, magnetic, and
thermodynamic properties. Reported magnetic ground states include ferromagnetic and antiferromagnetic, and in some cases even spin‑glass–like behavior. For example, ferromagnetic order is observed in Nd$_4$Ge$_7$ below 4 K, whereas $\alpha$-Sm$_3$Ge$_5$ and $\beta$-Sm$_3$Ge$_5$ show antiferromagnetic ordering below 30 K and 10 K, respectively.\cite{Bobev2004,Tobash2006,Zhang2013} 

The non-stoichiometric LnGe$_{2-x}$ compounds with 0$\leq x\leq$0.5, have been extensively investigated and are known to crystallize in more than twenty structure types including the hexagonal AlB$_2$ ($P$6/$mmm$) and  $P\bar{6}$2$c$, the tetragonal $\alpha$-ThSi$_2$ ($I$4$_1$/$amd$), and the orthorhombic $\alpha$-GdSi$_2$
($Cmcm$) and TbGe$_2$ ($Imma$).\cite{Tobash2006,Schobinger-Papamantellos1991} The structure type adopted within this family is strongly influenced by the size of the rare‑earth element and the concentration of Ge vacancies.\cite{Tobash2006,Zhang2013} The physical properties of the LnGe$_{2-x}$ compounds are similarly sensitive to subtle structural details and to the degree of disorder present.

According to previous studies, PrGe$_2$ shows the highest ferromagnetic temperature, $T_C$ = 19 K, when it adopts the tetragonal $\alpha$-ThSi$_2$ type structure\cite{Matthias1958,Sekizawa1966} in polycrystalline samples and $T_C$ = 17 K in single crystals grown by the Czochralski technique\cite{Boutarek1994}. However, a recent study on polycrystalline samples revealed that PrGe$_{2-x}$ orders ferromagnetically below $T_C$ = 14.6 K.\cite{Matsumoto2018} This study also assigned ThSi$_2$-type structure to their samples. A careful examination of the X-ray diffraction pattern in that work indicates the presence of Ge peaks and other unidentified small intensity peaks. Reference~\cite{Venturini1999a} lists five different phases, namely $\epsilon'$, $\alpha'$, $\beta'$, $\delta'$, and $\gamma'$, within 61–65 at.$\% $Ge in the Pr-Ge binary phase diagram. Since it is difficult to resolve and solve all structural intricacies with routine
X-ray analysis on polycrystalline samples, a single crystal study is preferred so that the directional dependence can be properly investigated without the averaging effect inherent to the polycrystalline samples.\cite{Tobash2007,Budko2014} \par

Given that PrGe$_{2-x}$ has a wide exposed liquidus line spanning from 62.5 to 84.5 at$\%$ Ge (the upper limit is an eutectic point),\cite{Gokhale1989} we report our extensive study on the crystal structure of PrGe$_{2-x}$ (Pr$_9$Ge$_{16}$ in this study) single crystals grown by the high-temperature flux technique. We establish that single crystals form in the
new $Fdd$2 structure type with ordered Ge vacancies over 77-84 at$\%$ of the nominal Ge concentration range used in flux growth. We report anisotropic physical properties and the $T$-$H$ phase diagram showing the suppression of the ferromagnetic order with the applied magnetic field along the easy crystallographic direction [010].\par

\section{Experimental Methods}

Single crystals of Pr$_9$Ge$_{16}$, in the form of plates, were grown from the high temperature binary solution rich in Ge.\cite{Canfield2009,Canfield1992} Pr pieces (Alpha Aesar, 99.5$\%$) and Ge pieces (Thermo Scientific, 99.999$\%$) in a molar ratio of 23:77 were loaded into the bottom alumina crucible of a 2 ml
Canfield crucible set\cite{Canfield2025,Canfield2016} which was then sealed inside a silica ampule under 0.25 atm of ultra-high-purity argon gas. The silica ampule was placed in a 50 ml alumina crucible and moved inside a muffle furnace. The growth was heated to 1463 K over 5 h, kept there for 0.5 h, and then cooled in
85 h to 1163 K at which temperature the excess molten Ge was decanted by centrifuging. The samples from this batch will be marked as Batch $\#$1. Two more batches were grown using the same starting composition, temperature profile, and the same furnace. The samples from these two batches will be marked as Batch $\#$2. Additional two batches with the starting compositions Pr$_{21}$Ge$_{79}$ and Pr$_{19}$Ge$_{81}$ were also grown at the same time in the same furnace, however, here the growths were cooled over 110 h to 1123 K, at which point the Ge flux was decanted in the centrifuge. The samples from these batches will be marked Batch $\#3$ and $\#4$, respectively.
Single crystals have a large "plate-like" morphology, inset of Fig.~\ref{PXRD}, and are stable in air.\par

Energy dispersive spectroscopy (EDS) was carried out using a VERSA 3D focused ion beam scanning electron microscope. Plate‑like single crystals of PrGe$_{2-x}$ were analyzed to verify the composition of the synthesized binary compound. The elemental ratios, obtained by averaging five EDS scans collected
from selected points on the crystals and normalizing to Pr, yielded a composition of Pr$_{9.0(3)}$Ge$_{14.9(3)}$; normalized to one Pr atom, this corresponds to PrGe$_{1.66}$. No additional elements were detected. 
The ratio of elements inferred from the EDS measurements is very close to the ratio Pr$_9$Ge$_{16}$ obtained from the crystallographic refinement of the crystal structure (see below). The small discrepancy in the Ge concentration arises from the undercounting of the lighter Ge atoms in the EDS measurements. We will use the composition Pr$_9$Ge$_{16}$ obtained from the single-crystal X-ray refinement throughout the manuscript.\par

Finely powdered single crystals were characterized at room temperature by powder X-ray diffraction (XRD) using Rigaku Miniflex 600/600-C benchtop X-ray diffractometer equipped with Cu K$_{\alpha}$ radiation. All data were collected with a 1 s exposure time and a 0.02$^\circ$ step size. Powder X-ray diffraction data were also reproducibly collected on multiple batches of ground single crystals at room temperature with the Bruker D2 Phaser diffractometer (Cu K$_\alpha$ radiation, $\lambda$ = 1.54184 \AA) operating at 30 kV/10 mA with a LYNXEYE XE-T detector. \par
 
The structure information of the crystals was investigated at Oak Ridge National Laboratory (ORNL) using a Rigaku Synergy-DW diffractometer equipped with a HyPix-ARC150 detector. A rotating molybdenum anode was used to generate X-rays with wavelength $\lambda$ = 0.71073 \AA. The samples were cooled by a Helium gas flow provided by an Oxford cryosystem with temperature controlled in the range of 90 - 300 K.
Data collection and reduction are achieved using Rigaku CrysAlisPro
software\cite{CrysAlisPro}. Single crystals with typical dimension of 0.02 $\times$ 0.05 $\times$ 0.05 mm$^3$  were selected and mounted using Paratone-N oil on MiTeGen micromount. The crystals were positioned at 50 mm from the detector and the data were collected for 2 hours for each piece. The single crystal X-ray data were collected on crystals from all four batches and showed a consistent chemical and structural composition. 

The crystal orientation was checked using the Laue Single Crystal Orientation System by Photonic Science, the plate-shape single crystal was placed along the X-ray beam direction and configured in the back-scattering geometry. The sample to detector distance was 50 mm, the Laue pattern can be indexed using the parent compound unit cell (weak reflections cannot be detected) with c-axis normal to the crystal plate, which corresponds to the $b$-axis of the Fdd2 of actual crystal structure.

Neutron diffraction was performed at the TOPAZ beamline at the Spallation Neutron Source, ORNL\cite{Schultz2014}. A single crystal was mounted to a magnetic-base loop pin and attached to an ambient goniometer equipped with two rotation stages and a nitrogen-gas-stream
temperature controller. After centering the sample and determining an initial unit cell, data-collection angles were selected with the CrystalPlan software, which optimizes the reciprocal space volume coverage\cite{Zikovsky2011}. The sample was cooled to 100 K and data were collected at 15 crystal orientations with 6 C (Coulomb) proton charge (about 1.25 hours) per angle. The diffraction measurements were merged into a reciprocal-space volume normalized to account for the incident spectrum, detector efficiency, and Lorentz corrections \cite{Michels-Clark2016,ARNOLD2014}.

Temperature and field-dependent magnetization data were collected on the cryogen free Quantum Design Physical Property Measurement System PPMS \textsuperscript{\textregistered} DynaCool\textsuperscript{\textregistered} using the Vibrating Sample Magnetometer (VSM) option. 

For transport measurements, plate-like samples of dimensions typically about $0.12 \times 0.12 \times 2$  mm$^3$ (resistivity) and $1 \times 0.12 \times 0.75$  mm$^3$ (Hall effect) were prepared by mechanical polishing. Four gold wire leads, in the standard four-probe configuration, were attached to each
sample using EpoTek{\textregistered} H20E silver epoxy.  An Oxford TeslatronPT, an integrated Cryofree\textsuperscript{\textregistered} superconducting magnet system (1.5-300 K and 12 T), was used to perform magneto-transport measurements. 

The single crystals from Batch $\#$1 were used for the neutron, magnetization, and transport measurements. The single crystal and powder XRD analysis were performed on the crystals from all batches grown and showed a consistent crystal structure and composition. \par

\section{Results and Discussion}
\subsection{Crystal Structure}
\subsubsection{X-ray Diffraction}

\begin{figure*}[tbh]
    \centering
    \includegraphics[width=0.75\linewidth]{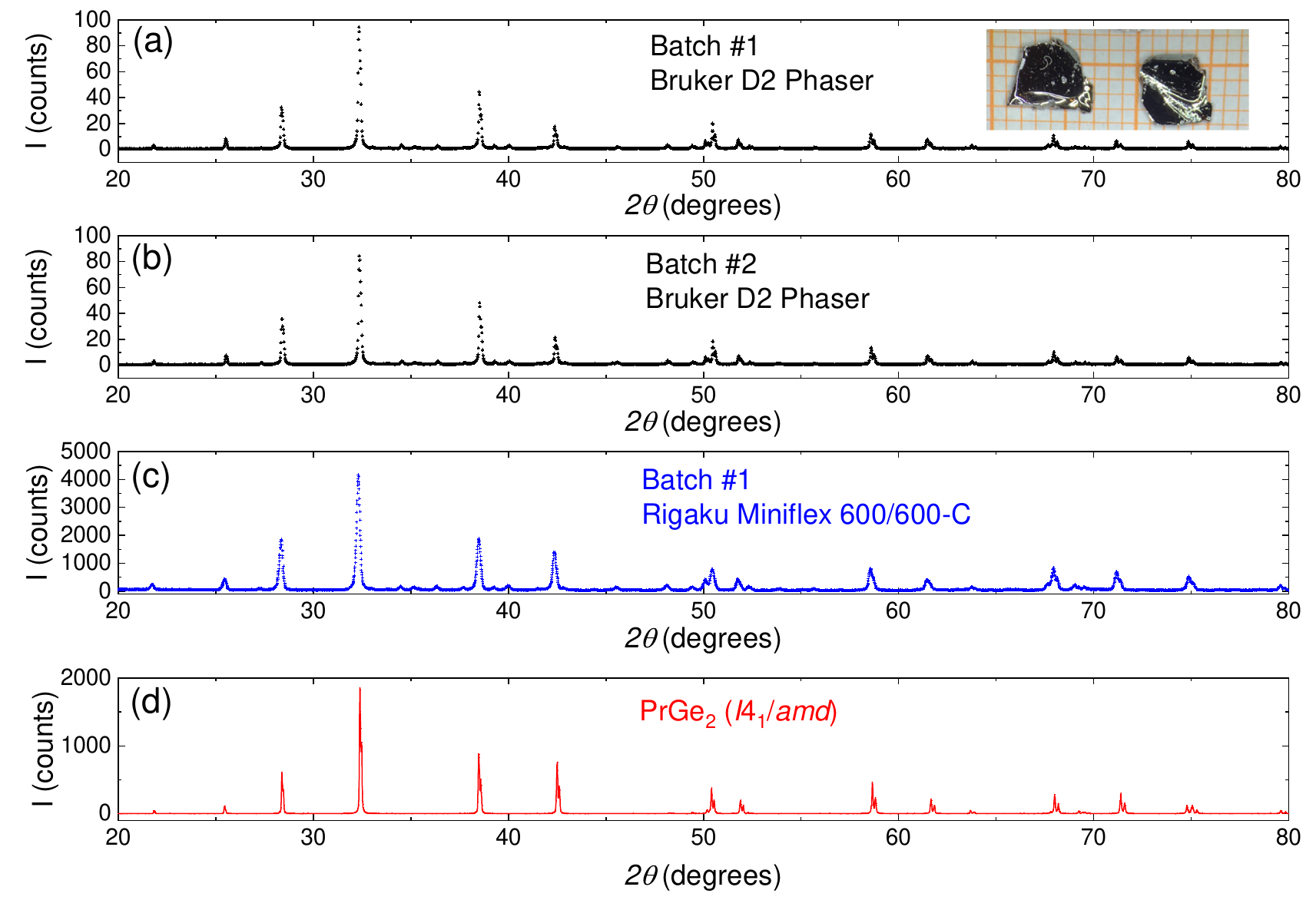}
    \caption{(a)-(c) X-ray diffraction patterns of Pr$_9$Ge$_{16}$ ground single crystals. The inset shows single crystals of Pr$_9$Ge$_{16}$. Squares in the background grid are 1 mm$^2$ each. (d) Simulated powder X-ray diffraction pattern for $I$4$_1$/$amd$ using crystallographic data from  Ref.~\cite{kozak2010structural}, $a$ = 4.251 and $c$ = 7.6131 \AA. The powder X-ray diffraction patterns in panels (a)-(c) clearly show consistent reflections that cannot be indexed by space group $I$4$_1$/$amd$.} \label{PXRD}
\end{figure*}
\begin{figure*}
    \centering
    \includegraphics[width=1\linewidth]{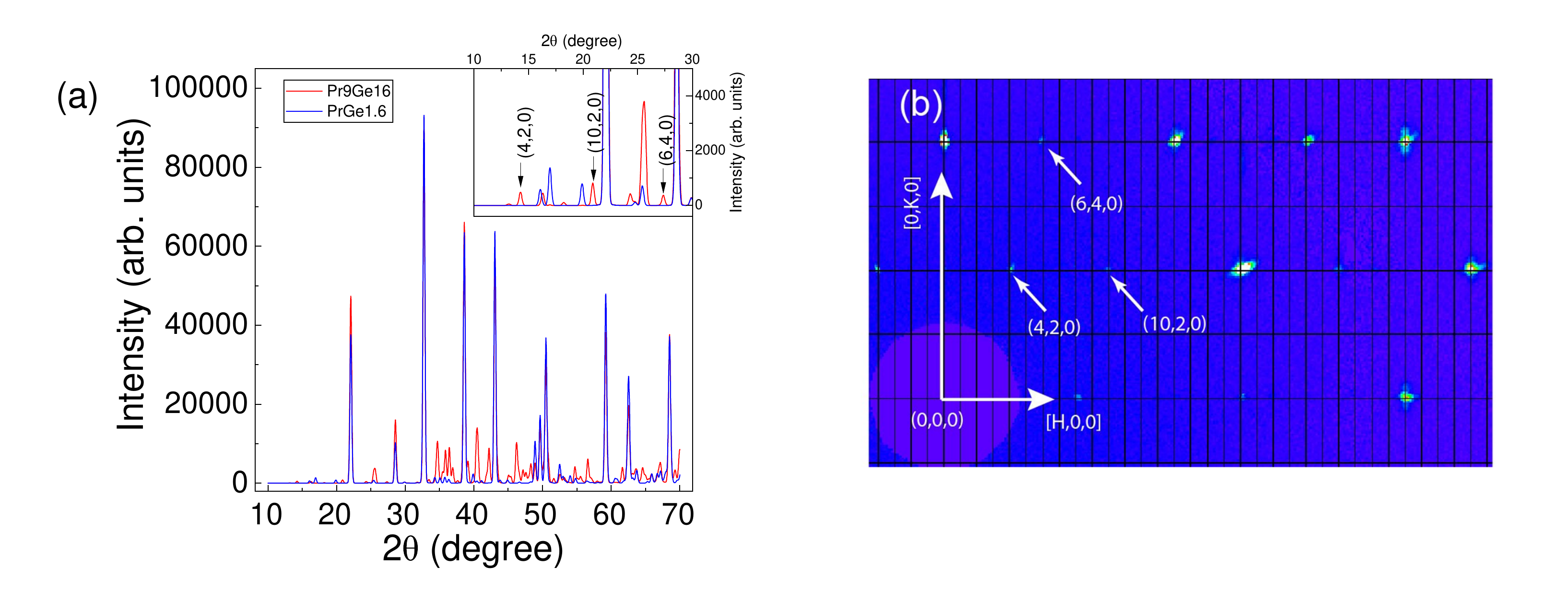}
    \caption{(a) Simulated neutron diffraction pattern for Pr$_9$Ge$_{16}$ and PrGe$_{1.6}$\cite{Schobinger-Papamantellos1990}. While the stronger intensity peaks coincide, the weaker peaks in Pr$_9$Ge$_{16}$, such as the (4,2,0), (10,2,0), and (6,4,0) reflections (indexed in our enlarged cell), are absent in the PrGe$_{1.6}$ structure. (b) The single crystal X-ray diffraction pattern of Pr$_9$Ge$_{16}$ in the ($h$,$k$,0) scattering plane. The weak reflections of (4,2,0), (10,2,0), and (6,4,0) are clearly visible, indicating the expanded unit cell.}
    \label{Fdd2}
\end{figure*}

\begin{figure}
    \centering
    \includegraphics[width=1\linewidth]{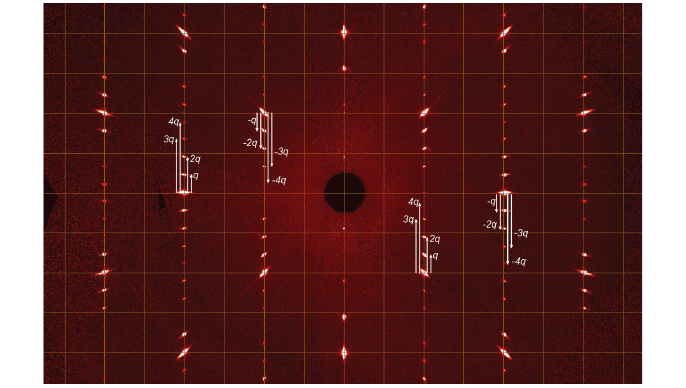}
    \caption{Single crystal X-ray diffraction in the ($h$,$k$,0) scattering plane showing satellite reflections up to the fourth order. The cell was indexed using the orthorhombic setting \cite{Venturini1999a} with $a$ = 6.0268, $b$ = 6.0347, and $c$ = 13.98~\AA.}
    \label{SR}
\end{figure}

\begin{table}
\caption{\label{tabl1} Crystallographic data determined through single-crystal X-ray diffraction. All data were collected at 250 K on Rigaku Synergy-DW diffractometer equipped with a HyPix-ARC150 detector. A rotating molybdenum anode was used to generate X-rays with wavelength $\lambda$ = 0.71073 \AA. }
\footnotesize\rm
\begin{tabular}{ll}
\hline
\hline
Chemical formula&Pr$_9$Ge$_{16}$\\
Space group&$Fdd$2 (No. 43)\\
%Structure type&&\\
\hline
$a$(\AA)&54.2345(9)\\
$b$(\AA)&13.9562(2)\\
$c$(\AA)&6.0298(1) \\
$V$({\AA}$^3$)&4564.00(13)\\
$\alpha$&90\textdegree \\
$\beta$&90\textdegree\\
$\gamma$&90\textdegree\\
Z&1\\
%Density (g/cm$^3$)&5.859&5.804\\
%absorption coefficient (mm$^{-1}$&27.543&27.017\\
Reflections collected &54041  \\
%\color{blue}Data/restraints/parameters&&\\
Goodness-of-fit on F$^2$&1.124 \\
R [$\rm F_o>4\sigma(F_o)$]    & 0.0336   \\
R [all data]& R$_1$=0.0430   \\
            & wR$_2$=0.0744  \\
%\color{blue}Largest diff. peak/hole / e$\mathrm{\AA}^{-3}$ &&  \\
Flack parameter & 0.37(5)  \\
\hline
\end{tabular}

\end{table}\begin{table}
\caption{\label{tabl2} Fractional Atomic Coordinates and Equivalent Isotropic Displacement Parameters ({\AA}$^2$) for Pr$_9$Ge$_{16}$. U$_{eq}$ is defined as 1/3 of of the trace of the orthogonalised U$_{ij}$ tensor.}
\footnotesize\rm
\begin{tabular}{llllll}
\hline
\multicolumn{6}{c}{Space group $Fdd$2}\\
\multicolumn{6}{c}{Chemical formula Pr$_9$Ge$_{16}$}\\\hline
Atom&\textit{x}&\textit{y}&\textit{z}&U$_{eq}$&Occ.\\
Pr1&0.36192(2)&0.74816(5)&1.13880(2)&0.0075(3)&1\\
Pr2&0.41577(2)&0.25475(6)&0.1353(2)&0.0077(3)&1\\
Pr3&0.44475(2)&-0.00406(4)&0.3725(3)&0.0064(2)&1\\
Pr4&0.47334(2)&0.25381(6)&0.6152(2)&0.0054(3)&1\\
Pr5&0.5&0&-0.1341(3)&0.0063(3)&1\\  
Ge1&0.38866(3)&0.4160(1)&0.8832(5)&0.0095(5)&1\\   
Ge2&0.47150(3)&0.3295(1)&0.1343(5)&0.0099(5)&1\\   
Ge3&0.44407(4)&0.4352(1)&0.3844(7)&0.0123(3)&1\\   
Ge4&0.41771(3)&0.3220(2)&0.6270(6)&0.0122(4)&1\\  
Ge5&0.39253(4)&0.5911(1)&0.9105(7)&0.0197(5)&1\\    
Ge6&0.50186(4)&0.4125(1)&-0.1345(6)&0.0124(4)&1\\  
Ge7&0.46548(4)&0.1515(1)&0.0672(8)&0.0235(5)&1\\ 
Ge8&0.42783(9)&0.1355(4)&0.7314(1)&0.107(2)&1\\ 
 \hline\hline
\end{tabular}
\end{table}

Powder X‑ray diffraction patterns for PrGe$_{2-x}$ single crystals from two batches are presented in Fig.~\ref{PXRD}(a)–(c). Comparison of these patterns with published diffraction data for the $I$4$_1$/$amd$, $Cmm$2, and $Imma$ space groups reported for PrGe$_{2-x}$ indicates that only the $I$4$_1$/$amd$ structure provides a satisfactory match, and even then only for the major intensity reflections. A simulated XRD pattern generated from the
crystallographic information data for the $I$4$_1$/$amd$ space group given in Ref.~\cite{kozak2010structural} is shown in Fig.~\ref{PXRD}(d). The additional low‑intensity peaks observed experimentally are not accounted for within the $I$4$_1$/$amd$ structural model. Similarly for the $Fdd2$ space group reported for PrGe$_{1.6}$ with unit cell size $a$ = 5.9193(3), $b$ = 17.7582(1), and $c$ = 14.128(1) \AA \cite{Schobinger-Papamantellos1990}, only the main-intensity reflections match, while the weaker reflections, for example, our (4,2,0), (10,2,0), and (6,4,0) peaks, are absent in that structure, see Fig.~\ref{Fdd2}. \par

Single‑crystal X‑ray diffraction measurements revealed that the crystals often contained multiple domain populations. Some crystals exhibited coexisting domains with modulated Ge‑atom displacements along both the $a$ and $b$ axes; others displayed a single dominant domain alongside a smaller secondary one; and in
some cases, only a single domain was present. Notably, even crystals smaller than 30 $\mu$m occasionally showed equally populated domains. These observations clearly indicate that the crystal structure has a symmetry lower than tetragonal.

\begin{figure}
    \centering
    \includegraphics[width=1\linewidth]{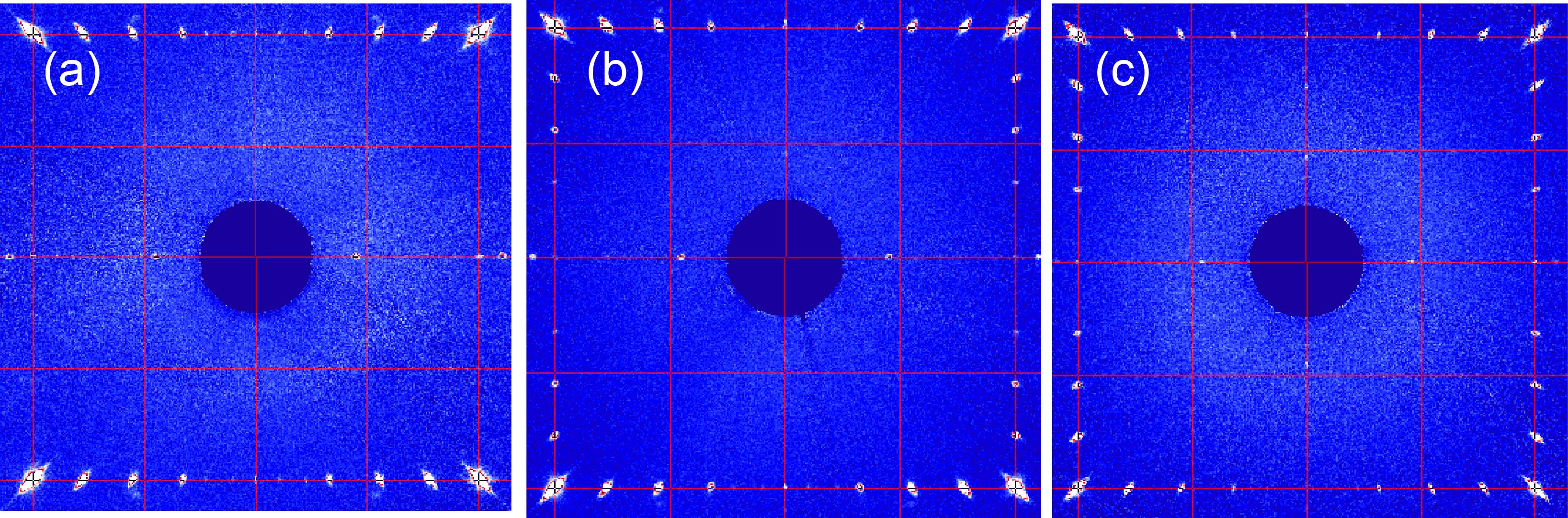}
    \caption{Single crystal X-ray diffraction patterns in selected samples in the ($h$,$k$,0) scattering plane showing different domain populations of (a) 100:0 (b) 70:30 and (c) 50:50 using the same orthorhombic setting as in Fig.~\ref{SR}.}
    \label{domain}
\end{figure}

If we consider sinusoidal modulation along either $a$ or $b$ axis, we will expect to observe only the first order satellite peaks in the X-ray diffraction pattern as a result of the Fourier transformation.\cite{Schobinger-Papamantellos1991} Given that
we observe up to the fourth order satellite reflections, with gradually reduced intensities with respect to the fundamental reflections, Fig.~\ref{SR}, the material has long-range order of Ge vacancy rather than sinusoidal modulation of Ge atoms' displacement. The superstructure is not consistent with a simple sinusoidal displacement modulation. Instead, the observation of higher-order satellite reflections up to fourth order, indicates a discrete vacancy-ordering distortion with substantial higher-order Fourier components, which gives rise to the observed harmonics. In addition, Fig.~\ref{domain} shows the projected diffraction patterns of selective single crystals with reflections indexed with the orthorhombic cell used in Ref.[\onlinecite{Venturini1999a}]. Note that the satellite peaks in panel (a) only appear along the $a$-axis, and become visible 
along  the $b$-axis [panel (b)], and finally reach equal intensity in panel (c). The evolution of the diffraction pattern clearly indicates that twinned domains are formed with different population ratios.

Since single crystal X-ray data were collected from multiple batches, the same satellite peaks with higher order harmonics and similar incommensurability $(0,\delta\sim0.45, 0)$ imply a stable crystal structure with robust vacancy order. The non-centrosymmetric space groups monoclinic $C$2 and orthorhombic $Fdd$2 were suggested for the possible structural solution by CrysAlisPro \cite{CrysAlisPro} for our system because they describe well the superstructure caused by the long range vacancy order. Another space group, orthorhombic $C$222$_1$, which
describes Pr$_4$Ge$_7$ well, ($C$222$_1$, Nd$_4$Ge$_7$ structure type with Ge vacancies order and a = 6.0282(9) \text{\AA}, b = 13.9690(16) \text{\AA}, and c = 12.048(2) \text{\AA}), \cite{Zhang2013,Venturini1999a,SHCHERBAN2009} does not work well for the given compound. The absolute structure cannot be reliably determined, with the residual electron density (16.71 $e$-\text{\AA}$^{-3}$) located at (0.3263, 0.3520, 0.3386) (0.59 \text{\AA} from Ge3). Attempts to model the germanium site disorder were unsuccessful. For example, modeling the residual electron density at the (0.3263, 0.3520, 0.3386) position as a split germanium site results in the residual electron density (14.20 $e$-\text{\AA}$^{-3}$) at the same position. The samples measured using both TOPAZ and laboratory X‑ray diffractometers consistently exhibit incommensurate Bragg peaks, and the data quality is sufficient to rule out the $C$222$_1$ structural model.

The crystallographic data obtained from the single‑crystal X‑ray diffraction refinements in the $Fdd$2 space group are summarized in Table I. Although the $Fdd$2 unit cell has twice the volume of the $C$2 cell, it contains fewer unique atomic positions—five unique Pr sites and eight unique Ge sites (Table II), and better describes the higher symmetry that is present in the arrangement of atomic sites, thus, is the most likely structure for Pr$_9$Ge$_{16}$. Averaging the incommensurate modulation vectors determined from 12 crystals yields $q$ = (0, 0.45, 0). Using the incommensurability relationship $q_y$=$n/p$, which corresponds to a chemical formula of Pr$_2$Ge$_{4-n/p}$, substituting $n/p$=0.45 gives  Pr$_2$Ge$_{3.55}$, which scales to the integer‑atom formula
Pr$_9$Ge$_{16}$. Although, the $Fdd2$ space group is non‑centrosymmetric and polar, the refined Flack parameter is not close to zero. Measurements collected from multiple crystals show that the Flack parameter varies from crystal to crystal due to inversion twinning, a common occurrence in such systems, where the atomic
arrangement is inverted within the unit cell while the $a$, $b$, and $c$ axes remain unchanged. It is important to emphasize that both the sample composition and the structural features observed across numerous crystals are highly consistent. Second‑harmonic‑generation measurements were not successful in confirming the lack of inversion symmetry, likely because the domain sizes — potentially on the order of microns — combined with the energetically degenerate inversion‑twin domains, reduce the sensitivity of this technique.

\begin{figure}
    \centering
    \includegraphics[width=1\linewidth]{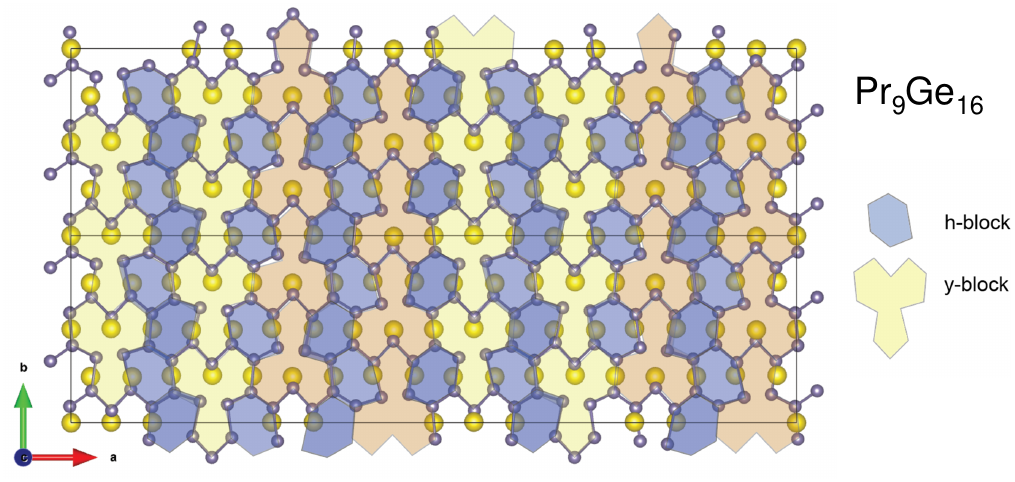}
    \caption{$h$-block and $y$-block schematics for $Fdd$2 space group for Pr$_9$Ge$_{16}$, with Pr and Ge ratio of 1.777.}
    \label{hy}
\end{figure}

To better visualize the origin of
the Ge vacancy in this structure, we adopt the cartoon representation of $y$ (Pr$_3$Ge$_2$) and $h$ (PrGe$_2$) building blocks from Ref.~\cite{Venturini1999a}. Undoped PrGe$_2$ consists solely of $h$‑blocks, whereas Pr$_3$Ge$_5$ contains only $y$‑blocks, and Pr$_4$Ge$_7$ corresponds to an equal mixture of the two. In the case of Pr$_9$Ge$_{16}$ (the composition
obtained for our crystals), the structure comprises a mixture of $y$‑ and $h$‑blocks, with a higer proportion of $h$‑blocks due to the higher Ge content, as shown in Fig.~\ref{hy}. Also, according to Ref.~\cite{Venturini1999a}, the incommensurability $q_y$ is expressed as $q_y=n/p$, where $n$ and $p$ are integers. The ratio of Ge and Pr can be expressed by (4-$n/p$)/2, in our case, $n/p$=0.445, so the ratio is 1.7775, close to the chemical formula Pr$_9$Ge$_{16}$ (16/9)=1.777. 

\begin{figure*}[htp]
    \centering
    \includegraphics[width=0.8\linewidth]{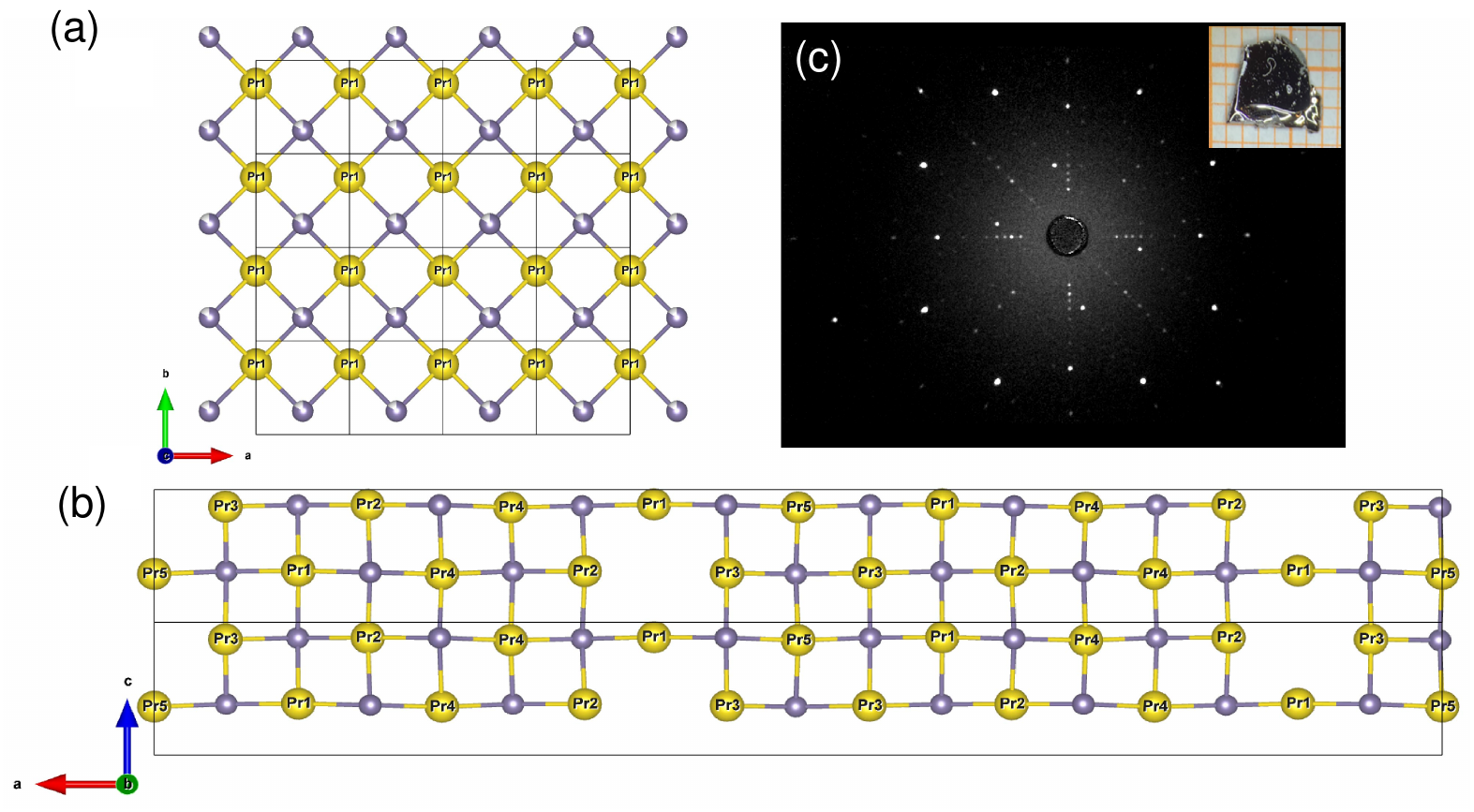}
    \caption{Projections of unit cells for (a) the averaged $I$4$_1$/$amd$ and (b) refined $Fdd$2 space groups. (c) X-ray Laue diffraction pattern of Pr$_9$Ge$_{18}$ single crystal along [010] direction. The $b$ axis is perpendicular to the the plane of the naturally grown single crystal shown in the inset.}
    \label{CS}
\end{figure*}

Figure~\ref{CS} illustrates the comparison between the refined $Fdd$2 structure and the $I$4$_1$/$amd$ structure. In the original smallest tetragonal cell ($I$4$_1$/$amd$), the Pr atoms form a square lattice, and the average Ge atom occupies the center of the square lattice, Fig.~\ref{CS}(a). However, in the $Fdd$2 structure, Fig.~\ref{CS}(b), the Pr atoms still form an approximately square‑lattice arrangement, whereas the Ge sublattice is significantly distorted. Several of
the square‑lattice centers lack Ge atoms entirely (i.e., Ge vacancies). This refined structure is distinct from previously reported Pr–Ge phases and represents a new crystal structure at the highest Ge‑rich end of the Pr–Ge binary phase diagram. 

Figure~\ref{CS}(c) presents an X-ray Laue diffraction pattern along the [010] direction, which is the direction perpendicular to the naturally formed plane of single crystals.

\subsubsection{Neutron Diffraction}
\begin{figure}
    \centering
    \includegraphics[width=1\linewidth]{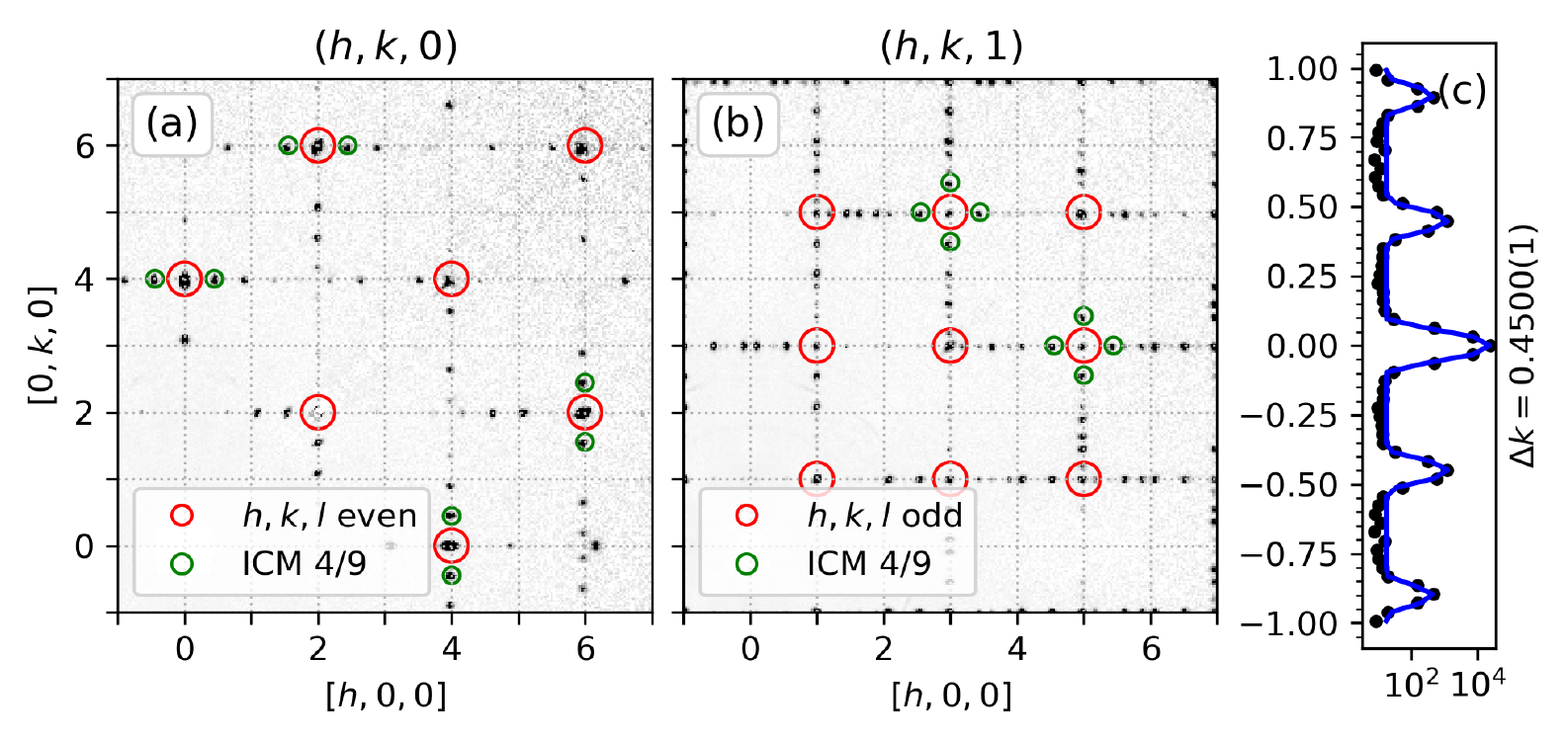}
    \caption{Neutron diffraction of Pr$_9$Ge$_{16}$ single crystal measured at $T$ = 100 K indexed in non-conventional $F$4$_1$/$ddm$ space group with tetragonal lattice parameters $a$ = 5.974(6) and $c$ = 13.99(3) \AA. Selected reciprocal-space slices with normal along the [0,0,$l$] direction, (a) ($h$,$k$,0)and (b) ($h$,$k$,$1$), show contributions from two non-merohedral twin domains related by a 90$^{\circ}$ rotation about the $c$-axis. Integer Bragg reflections satisfying the face-centering conditions are highlighted in red, while incommensurate satellite reflections are indicated in green. The satellites are displaced along the $k$-direction by a modulation wavevector component $\Delta k$ that is close to the composition-derived rational value 4/9. (c) A fit to a representative line cut along the [0,$k$,0] direction near the (4,0,0) reflection yields a refined value $\Delta k$ = 0.4500(1); additional features at approximately integer multiples of this value indicate the presence of higher-order harmonic satellites.}
   
    \label{neutron}
\end{figure}

Neutron diffraction measurements of Pr$_9$Ge$_{16}$ single crystal collected at $T$=100 K reveal additional reflections beyond those expected for the parent
ThSi$_2$-type structure \cite{Venturini1999a}. The data are well-indexed in the non-conventional $F$4$_1$/$ddm$ setting with tetragonal lattice parameters $a$ = 5.974(6)~\r{AA} 
and $c$ = 13.99(3)~\r{AA}, Fig.~\ref{neutron}. 
This space group was used to simplify the description of the data because the larger sample used for neutron diffraction contains domains. Reciprocal-space reconstructions show the presence of two non-merohedral twin domains related by a 90° rotation about
the $c$-axis, producing two interpenetrating sets of Bragg reflections in the ($h$,$k$,$l$) planes \cite{Parsons2003}. After accounting for the twin relationship, additional satellite reflections are observed around the fundamental Bragg peaks, offset along the $k$-direction. These satellites violate the face-centering conditions of the average structure and are therefore
attributed to an incommensurate modulation that can be related to the Ge vacancies. Line cuts taken along the [0,$k$,0] direction near strong parent reflections, such as (4,0,0), resolve a fundamental modulation wavevector component $\Delta k$ = 0.4500(1). In addition to the first-order satellites, additional reflections are detected at approximately integer multiples of this
displacement, indicating the presence of higher-order harmonic satellites. The presence of multiple harmonic orders confirms a strongly non‑sinusoidal modulation, consistent with long‑range ordering of Ge vacancies rather than a simple displacive distortion, as also indicated by the single‑crystal X‑ray diffraction results. The refined value lies close to the composition-derived rational value 4/9, supporting an interpretation in which the incommensurate modulation reflects vacancy ordering within the ThSi$_2$-derived framework of Pr$_9$Ge$_{16}$ \cite{Venturini1999a}.

\subsection{Magnetization}

\begin{figure}
    \centering
    \includegraphics[width=1\linewidth]{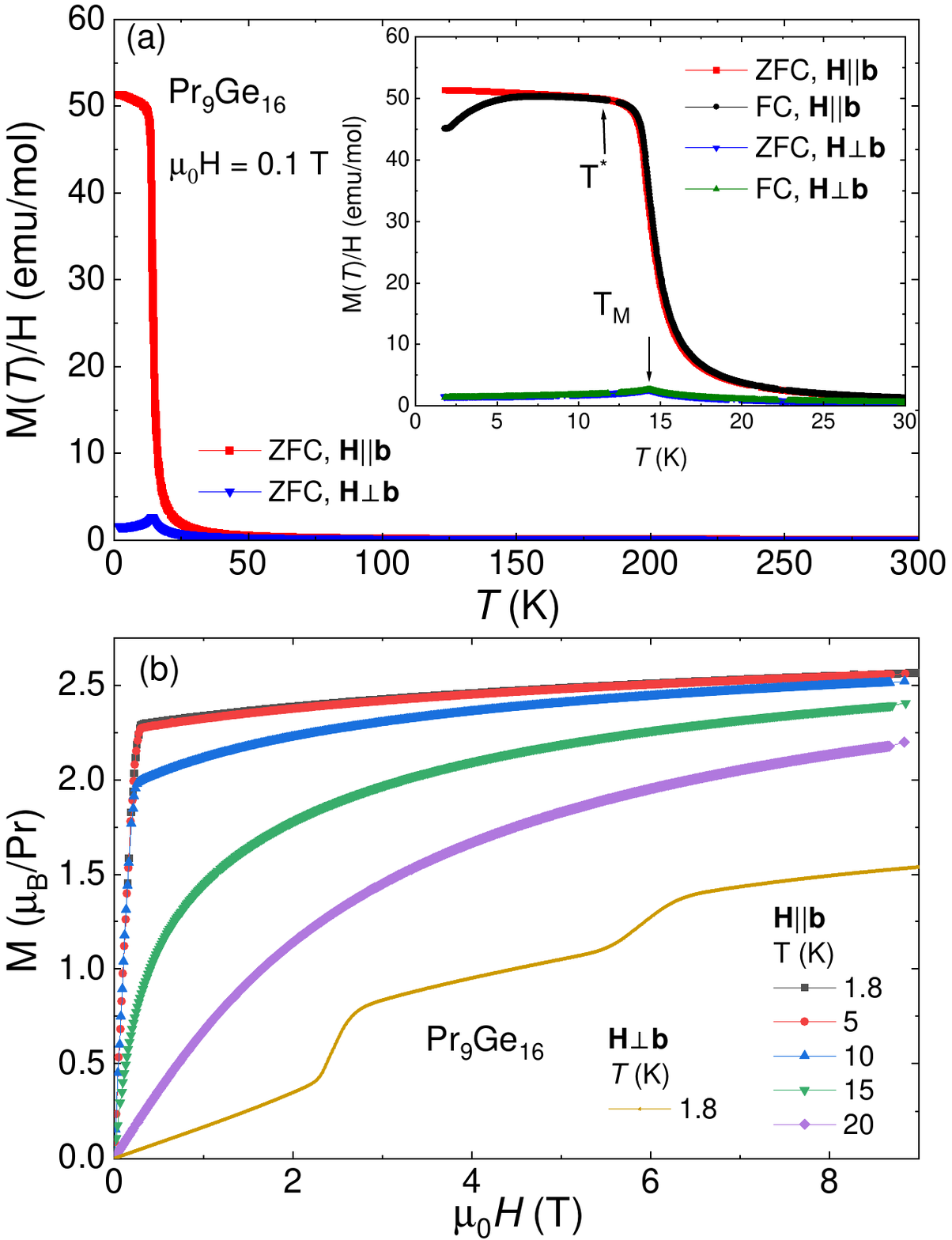}
    \caption{(a) Temperature-dependent magnetization $M$($T$)/$H$ of Pr$_9$Ge$_{16}$ single crystal measured under zero-field-cooled (ZFC) conditions in an applied magnetic field of 0.1 T, with the magnetic field applied perpendicular to (\textbf{H}$\parallel$\textbf{b}) and parallel to (\textbf{H}$\bot$\textbf{b} ) the plate-like surface of the single crystal. The inset displays temperature-dependent magnetization $M$($T$)/$H$ of Pr$_9$Ge$_{16}$ measured under both ZFC and field-cooled
    (FC) conditions in an applied magnetic field of 0.1 T. T$^*$ marked by the arrow is the temperature below which ZFC and FC curves for \textbf{H}$\bot$plate diverge. (b) Field-dependent magnetization data $M(H)$ of Pr$_9$Ge$_{16}$ at 1.8, 5, 10, 15 and 20 K for \textbf{H}$\|$\textbf{b} and 1.8 K for \textbf{H}$\bot$\textbf{b}.}
    \label{MTH}
\end{figure}

Temperature-dependent magnetization $M$($T$)/$H$ data, which were collected under zero-field cooled (ZFC) and field cooled (FC) conditions, are shown in Fig. \ref{MTH}(a). Anisotropy is evident between the data for \textbf{H}$\|$\textbf{b} and \textbf{H}$\bot$\textbf{b}. The anisotropy ratio at T$\sim$1.8 K is $\sim$37.4 implying the existence of a ferromagnetic order along
the direction perpendicular to the plane of the single crystal and defines the out of plane axis, $b$-axis as an easy axis. The bifurcation of the ZFC and FC data at $\sim$ 10 K is consistent with the irreversible behavior while cooling the sample in the magnetic field into the ferromagnetic state. The sharp cusp at $T_M \sim$ 14.2 K in the ZFC and FC curves for \textbf{H}$\bot$\textbf{b} indicates an antiferromagnetic-like transition for this field orientation. As shown in the inset of Fig. \ref{MTH}(a), the FC and ZFC data for \textbf{H}$\bot$\textbf{b} are essentially the same for all temperatures considered. However, a small hysteresis in ZFC and FC data in the ordered state is observable as well. A similar behavior is observed in the high temperature data for \textbf{H}$\|$\textbf{b}. Below $T^*$ $\sim$ 12.4 K, FC and ZFC data deviate so
that the $M(T)/H$ values for the FC data are lower than those for the ZFC data. This indicates an irreversible magnetic behavior that could be due to the ferromagnetic phase transition (formation of ferromagnetic domains) at this critical temperature. The ZFC data from 150 – 300 K were fitted using the Curie-Weiss law, $\chi$ = $C$⁄ ($T$-$\theta$$_p$), where $\chi$ is the magnetic
susceptibility, $C$ is the Curie constant and $\theta$$_p$ is the Weiss constant. This resulted in $\theta$$_p$ $\cong$ 32.77 K and $\theta$$_p$ $\cong$ 1.17 K for \textbf{H}$\|$\textbf{b} and \textbf{H}$\bot$\textbf{b}, respectively. According to Callen's theory \cite{Callen1960a,Callen1960b,Callen1961a,Callen1961b,Armstrong2013}, large magnetocrystalline anisotropy leads to a strongly anisotropic $\theta_p$. This can explain the difference observed in $\theta_p$. In addition, the calculated effective magnetic moments for \textbf{H}$\|$\textbf{b} and \textbf{H}$\bot$\textbf{b} are $\mu _{eff}$ = 3.38 $\pm$ 0.11  $\mu _B$ ⁄ Pr$^{3+}$  and $\mu _{eff}$ = 3.54 $\pm$ 0.11  $\mu _B$ ⁄ Pr$^{3+}$, respectively. The polycrystalline average, $\chi$ = (2$\chi _{H\perp b}$ + $\chi
_{H\| b}$ )⁄3 resulted in $\theta _p ^{poly}$ $\cong$ 13.65 $\pm$ 0.03 K and $\mu _{eff}^{poly}$ = 3.48 $\pm$ 0.11 $\mu _B$ ⁄ Pr$^{3+}$ which is close to 3.58 $\mu _B$, the free-ion value of Pr$^{3+}$. \par

Figure \ref{MTH}(b) displays the field dependence of the magnetization $M$($H$) data measured at constant temperatures for the magnetic field applied perpendicular and parallel to the crystal plane. For \textbf{H}$\|$\textbf{b}, the $M$($H$) data at 1.8, 5, and 10 K show a spin-flip transition below 0.3 T. The data set for 15 K, above the ferromagnetic transition, does not have any features and is similar to the Brillouin function. The $M$($H$) data at 1.8 K for \textbf{H}$\bot$\textbf{b} shows two metamagnetic
transitions with a value of $M$ reaching 1.5 $\mu_B$ at 9 T. At this magnetic field, the anisotropy between in-plane and out of plane magnetization is much smaller than at 0.1 T. According to the magnetization isotherms, the saturation magnetization ($M_{sat}$) values do not reach the value for the free Pr$^{3+}$ ion, which is 3.20 $\mu _B$ per Pr$^{3+}$ ion.
These discrepancies in the resulting values can be due to the effects of the Crystal Electric Field (CEF)\cite{Tarasov2022}. The Ge vacancies in Pr$_9$Ge$_{16}$ can also influence magnetic properties.\cite{Gondek2004} \par

\subsection{Transport}

\begin{figure}
    \centering
    \includegraphics[width=1\linewidth]{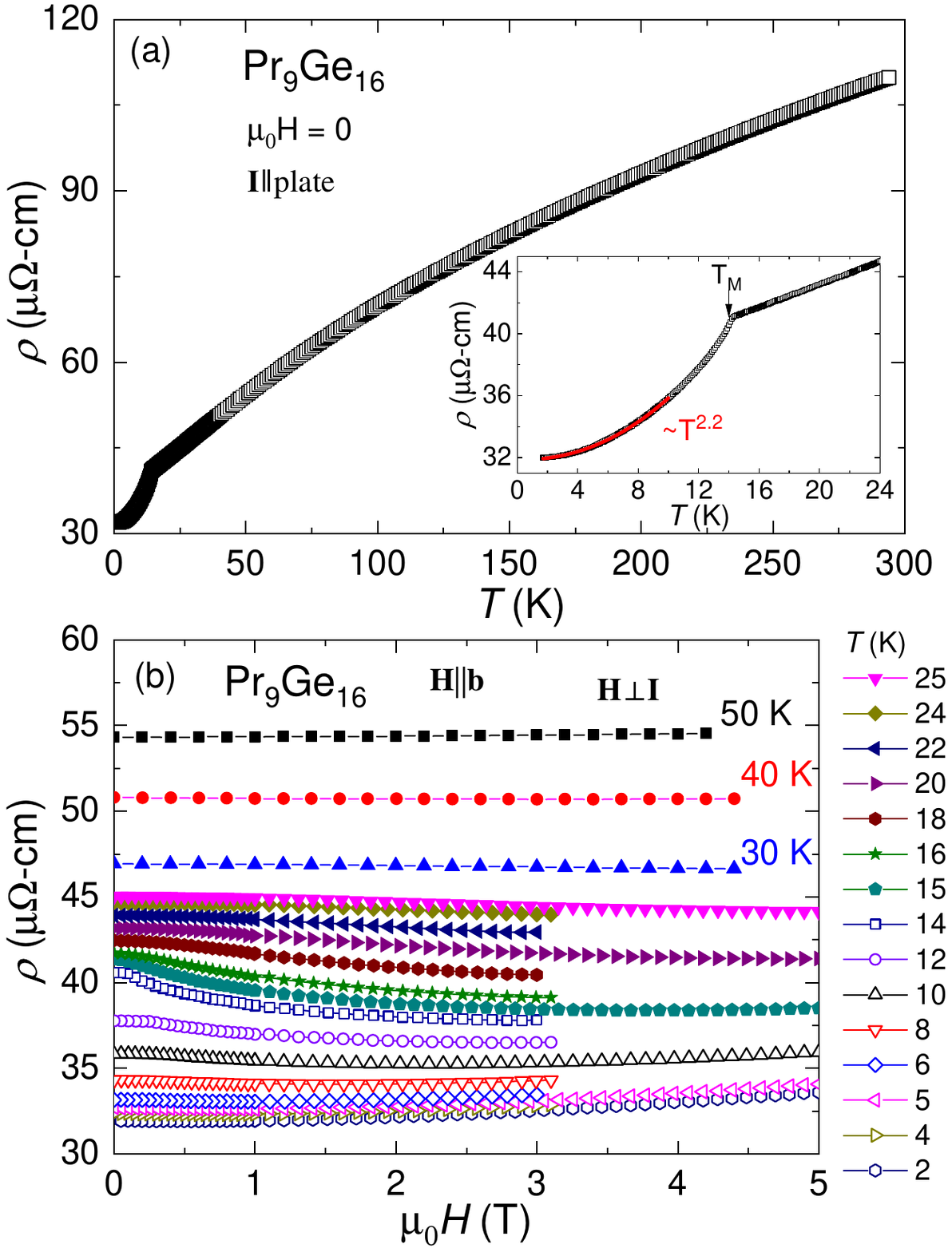}
    \caption{(a) Zero-field, temperature-dependent resistivity $\rho$(T) data of Pr$_9$Ge$_{16}$ single crystal. The lower right inset displays the low temperature section of the $\rho$(T) data with the magnetic transition marked with an arrow. The red line is the fit to the power law $\rho$
    =$\rho_0$ + A$\times T^n$. (b) Field-dependent resistivity for
    \textbf{H}$\|$\textbf{b} for Pr$_9$Ge$_{16}$ single crystal.}
    \label{RTH}
\end{figure}

\begin{figure*}
    \centering
    \includegraphics[width=0.8\linewidth]{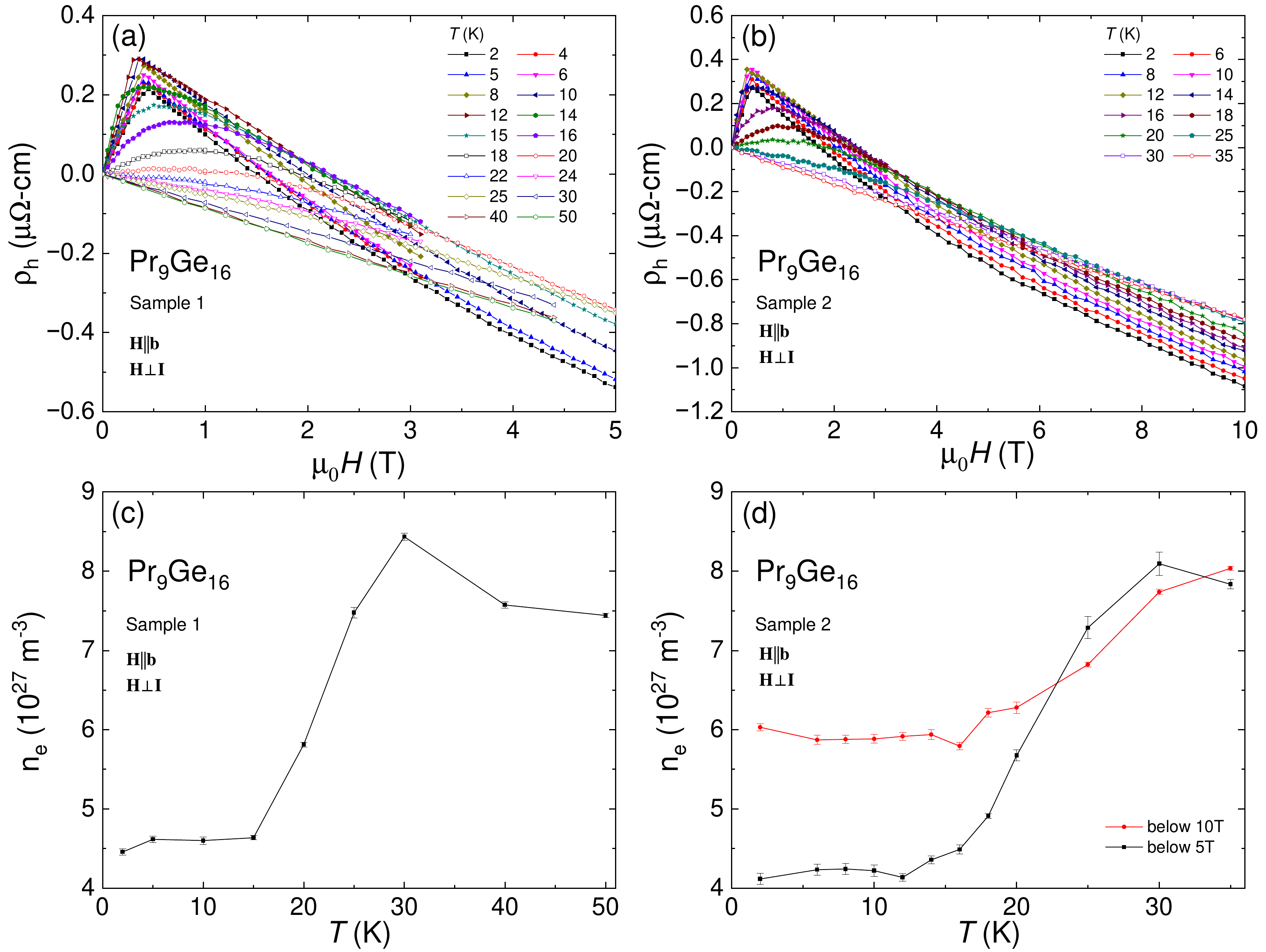}
    \caption{Field dependent Hall resistivity $\rho_h$ isotherms for (a) sample 1 (measured up to 5 T) and (b) sample 2 (measured up to 10 T), \textbf{H}$\|$\textbf{b}. Carrier concentration plotted as a function of temperature for (c) sample 1 and (d) sample 2. Carrier concentrations were extracted by fitting the linear
    portion of $\rho_{xy}$ curves using one band model at high field in the available data range. In panel (d), the black data set corresponds to the carrier concentration calculated for the same magnetic field range as in (c).} \label{RHall}
\end{figure*}

The temperature-dependent resistivity data of Pr$_9$Ge$_{16}$ show metallic behavior, Fig. \ref{RTH}(a). The residual resistivity ratio (RRR) is $\rho$(294 K)⁄$\rho$($\sim$1.7 K)  $\approx$ 3. Because flux-grown single crystals show good crystallinity, this low RRR could be due to disorder caused by Ge deficiencies or carrier compensation. Some materials, such as PrAlGe \cite{Puphal2019,Destraz2020}, also show a low RRR=2 without substantial structural disorder. A kink, associated with the magnetic transition, is observed at $T_M$ = 14.3 $\pm$ 0.1 K, similar with that observed for polycrystalline PrGe$_{2-x}$ samples\cite{Matsumoto2018}. Below the transition temperature in
the region 1.7–10 K, the resistivity behaves as $\rho$ =$\rho_0$ + A$\times T^{2.2}$ as shown in the lower right inset of Fig. \ref{RTH}(a). This power constant is nearly quadratic, as expected for the Fermi liquid, which typically exhibits an electrical resistivity that varies with the square of the temperature. From the transition temperature up to $\sim$ 40 K, the
resistivity increases linearly after which a sublinear behavior is observed. This deviation from linearity could be due to the CEF effects.\par 

Figure \ref{RTH}(b) shows the field-dependent resistivity data $\rho(H)$ collected at constant temperatures with \textbf{H}$\|$\textbf{b}. The magnetoresistance is very small for all temperatures measured. Below 12 K, in the ordered state, magnetoresistance has both negative and positive contributions. However, above 12 K, magnetoresistance
remains negative throughout the measured magnetic field range. \par

Figures \ref{RHall}(a) and \ref{RHall}(b) display Hall resistivity $\rho_{xy}$ data as a function of the magnetic field for samples 1 and 2, respectively. Both samples show qualitatively similar functional behavior below 5 T. The small inconsistency in the value of $\rho_{xy}(H)$ data could be due to geometric errors or a slight
misalignment of the sample during mechanical polishing. It should be noted that the single crystals form as plates; however, the outer layer is not completely flat. This may also contribute to a slight misalignment of the sample during mechanical polishing. In addition, since the crystals' habit is such that it is not possible to visually
ascertain the in-plane crystallographic directions, misalignment of the applied current with respect to the in-plane crystallographic axes can also contribute to some differences in the measured value of the Hall resistivity of the two samples. Because the fit of the field-dependent Hall resistivity data to the two-band model did not converge, we fitted the data to a one-band model to extract the carrier concentration. The results of the fit of the one band model are shown in Figs. \ref{RHall} (c) and (d) for the two samples and indicate that the Hall coefficient is negative. It should be noted that the majority of the carriers are electrons of the order of 10$^{27}$ m$^{-3}$, similar to what was observed for CeAlGe\cite{Drucker2023}. The carrier
concentration is constant up to the ferromagnetic transition temperature, at which point it doubles and has a maximum at $\sim$ 30 K. This behavior is consistent between the two samples and may point to the Fermi surface reconstruction at the ferromagnetic transition temperature. \par

\begin{figure}
    \centering
    \includegraphics[width=1\linewidth]{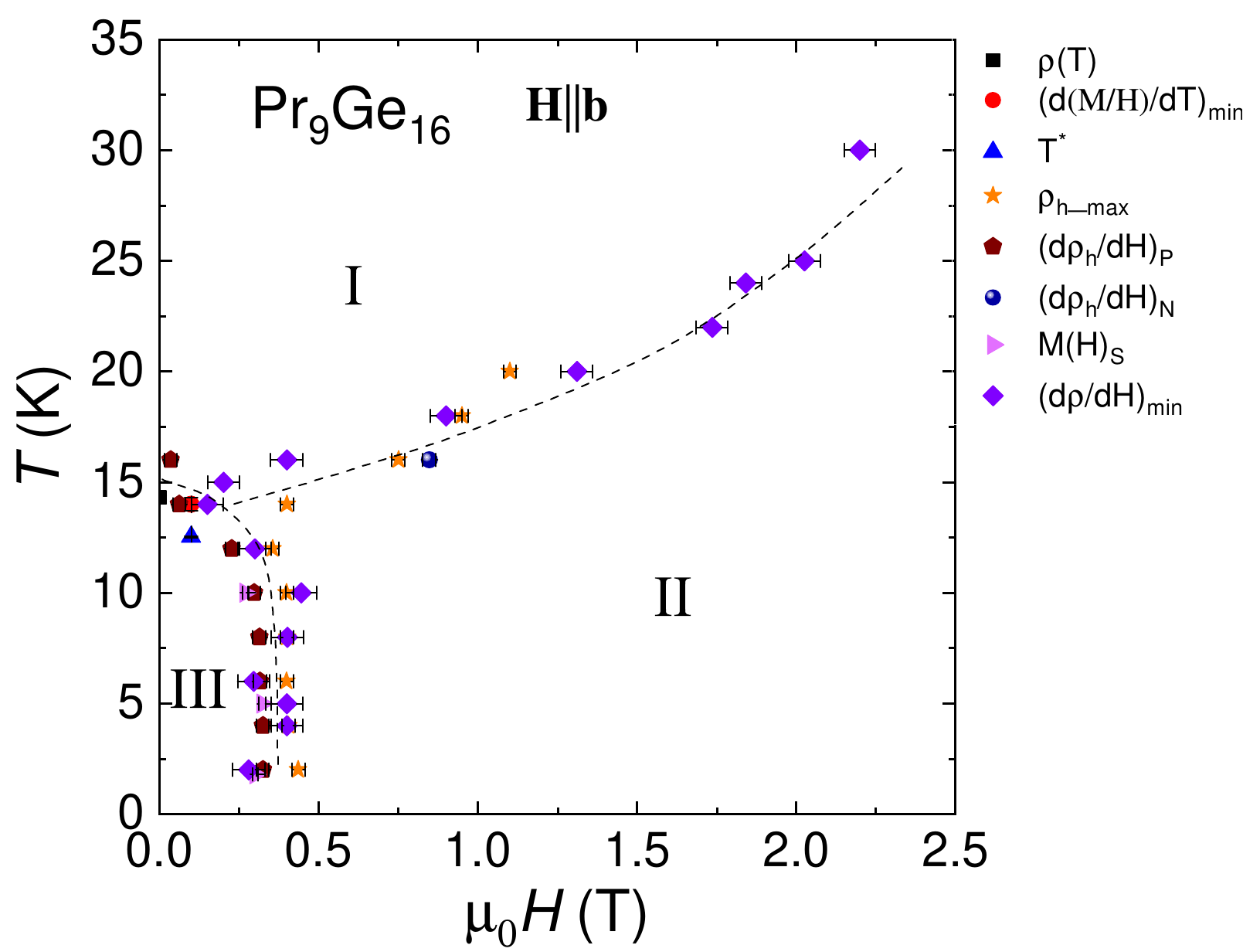}
    \caption{$T$-$H$ phase diagram of Pr$_9$Ge$_{16}$ single crystals for \textbf{H}$\|$\textbf{b}. Phase region I corresponds to paramagnetic (PM) phase, region II is field-induced ferromagnetic, and III corresponds to long range magnetic order state. Dashed lines are guides to the eye.}
    \label{HT}
\end{figure}

\subsection{$T$-$H$ phase diagram}
Figure \ref{HT} shows the $T$-$H$ phase diagram of Pr$_9$Ge$_{16}$ for \textbf{H}$\|$\textbf{b} based on the features observed in the magnetization, susceptibility, longitudinal resistivity, and Hall resistivity measurements as the legends indicate. The phase diagram illustrates the presence of three phases: paramagnetic (regions I), field-induced ferromagnetic (II) and region III that corresponds to the long-range magnetic order. The long range magnetic order, based on the ZFC and FC magnetization data, is of ferromagnetic nature. Neutron diffraction studies at low temperatures are needed to elucidate the details of the nature of long range order in phase III. The phase diagram indicates that a relatively small magnetic field of $\sim$0.4 T at 2 K is sufficient to suppress the magnetic order. The $T$-$H$ phase diagram is surprisingly very simple given the complex crystal structure and five distinct Pr crystallographic sites. In contrast, three transitions were observed in single crystals of the related compound CeGe$_{1.76}$\cite{Budko2014} and the subsequent neutron study\cite{Jayasekara2014} revealed a complex series of magnetic transitions with transitions of incommensurate to commensurate to incommensurate order. As a result, very complex $T$-$H$ phase diagrams were obtained. 

Pr ion is a non-Kramers ion. Pr$_9$Ge$_{16}$ has five unique Pr and eight unique Ge crystallographic sites resulting in 200  symmetry-related atoms within the unit cell. Four of the five unique Pr crystallographic sites have site symmetry of 1, while the fifth Pr site has symmetry of 2. Under these low site symmetry environments, the CEF splits the nine-fold degenarate (J=4) ground state multiplet of Pr$^{3+}$ into nine distinct non-degenerate singlet states, resulting in a complex CEF level scheme. The magnitude of the splitting depends on the local environment of the Pr ion. Because the ground sate is a singlet, no magnetic order is expected. However, we do observe magnetic order below 14.3 K. This suggests that the two lowest singlet levels are separated by a sufficiently small CEF splitting to form a pseudo-doublet ground sate, which gives rise to the magnetic order. The low-site symmetries present considerable challenges for studying the energy levels and CEF level splitting of Pr ion in crystals, because assigning spectral features to irreducible representations is difficult and may be unreliable. To determine the precise CEF level scheme of Pr$_9$Ge$_{16}$, inelastic neutron scattering experiments are needed.

The $T$-$H$ phase diagram of Pr$_9$Ge$_{16}$ is also very similar, in terms of the critical magnetic field needed to suppress the FM order and the mapped phases, to the one delineated for PrAlGe\cite{Destraz2020}. Although the temperature-dependent magnetization and field-dependent magnetization for the easy axis are also very similar to those of PrAlGe, transport properties, particularly the Hall resistivity, are very different from those of PrAlGe. PrAlGe forms in the $I$4$_1md$ space group, which is an ordered variant of the $I$4$_1$/$amd$ structure.\cite{Hodovanets2018} While PrAlGe hosts Weyl fermions\cite{Sanchez2020} and the anomalous Hall effect (AHE) was observed\cite{Destraz2020}, Pr$_9$Ge$_{16}$ does not show any measurable AHE. Future band structure calculations and ARPES measurements would be needed to verify if Pr$_9$Ge$_{16}$ can host Weyl fermions.

\section{Conclusion}
In summary, we discovered a new crystal structure on the Ge‑rich side of the Pr–Ge binary phase diagram. Using a high‑temperature flux technique, we grew single crystals of Pr$_9$Ge$_{16}$, which adopt a previously unreported $Fdd$2 structure type with ordered Ge vacancies. From magnetization measurements, we identified the crystallographic $b$ axis perpendicular to the crystal plane as the easy axis of magnetization. Temperature‐-dependent resistivity measurements reveal metallic behavior with a distinct anomaly at $T_C$ = 14.3 K. Hall resistivity data indicate that electron-like carriers
dominate, with a carrier concentration on the order of 10$^{27}$ m$^{-3}$. The magnetic order is readily suppressed by a magnetic field of approximately 0.4 T applied along the easy $b$ axis.

\section{Acknowledgments}

The authors thank D. Mandrus for fruitful discussions. J.S.T.W, and H.H. acknowledge support from the US National Science Foundation Division of Materials Research Award DMR-2316869. Magnetization measurement performed at the University of Arizona was supported by the U.S. Department of Energy (DOE), Office of Science,
Basic Energy Sciences (BES) under Award DE-SC0025301. B.R.B acknowledges support from the US National Science Foundation under Award DGE-2137419. A portion of this research used resources at the Spallation Neutron Source, a DOE Office of Science User Facility operated by the Oak Ridge National Laboratory. The beam
time was allocated to TOPAZ on proposal number IPTS-34800.1. The authors acknowledge Alexis Dominguez Montero and Christina Hoffmann for the initial collection of neutron diffraction data at TOPAZ. J.Y.C. acknowledges Welch Foundation AA-2056-20240404 and M.G.A. acknowledges DE-SC0022854 for partial support of this project. The SHG work by S.G. and L.W.  was supported by the US Office of Naval Research through the grant N00014-24-1-2064.
\newpage

\bibliography{bibiliography}

\end{document}